\documentclass[aps,prl,twocolumn,superscriptaddress,a4paper,english,longbibliography]{revtex4-2}
\usepackage{times}
\usepackage{color}
\usepackage[colorlinks=true,urlcolor=blue,citecolor=blue]{hyperref}
\usepackage{url}
\usepackage{breakurl}
\usepackage{soul}
\usepackage{graphicx}
\usepackage{amsmath}
\usepackage{subfigure}
\usepackage{xcolor}
\usepackage{amssymb}
\usepackage{latexsym}
\usepackage{hyperref}
\DeclareGraphicsExtensions{.png,.eps}
\usepackage{float}
\usepackage{makecell}
\usepackage{diagbox}
\usepackage{setspace}
\usepackage{appendix}
\usepackage{etoolbox}

\newcommand {\R}{\textcolor {black}}
\newcommand {\RR}{\textcolor {black}}
\usepackage{xcolor}
\usepackage[urlcolor=blue]{hyperref}      
\hypersetup{
	colorlinks = true,                    
	citecolor = {blue},
	linkcolor = {purple},
}

\begin{document}	
	
\title{Tensor Network Efficiently Representing Schmidt Decomposition of Quantum Many-Body States}

\author{Peng-Fei Zhou}
\affiliation{Center for Quantum Physics and Intelligent Sciences, Department of Physics, Capital Normal University, Beijing 10048, China}
\author{Ying Lu}
\affiliation{Center for Quantum Physics and Intelligent Sciences, Department of Physics, Capital Normal University, Beijing 10048, China}
\author{Jia-Hao Wang}
\affiliation{Center for Quantum Physics and Intelligent Sciences, Department of Physics, Capital Normal University, Beijing 10048, China}
\author{Shi-Ju Ran} \email[Corresponding author. Email: ] {sjran@cnu.edu.cn}
\affiliation{Center for Quantum Physics and Intelligent Sciences, Department of Physics, Capital Normal University, Beijing 10048, China}
\date{\today}

\begin{abstract}
    Efficient methods to access the entanglement of a quantum many-body state, where the complexity generally scales exponentially with the system size $N$, have long a concern. Here we propose the Schmidt tensor network state (Schmidt TNS) that efficiently represents the Schmidt decomposition of finite- and even infinite-size quantum states with nontrivial bipartition boundary. The key idea is to represent the Schmidt coefficients (i.e., entanglement spectrum) and transformations in the decomposition to tensor networks (TNs) with linearly-scaled complexity versus $N$. Specifically, the transformations are written as the TNs formed by local unitary tensors, and the Schmidt coefficients are encoded in a positive-definite matrix product state (MPS). Translational invariance can be imposed on the TNs and MPS for the infinite-size cases. The validity of Schmidt TNS is demonstrated by simulating the ground state of the quasi-one-dimensional spin model with geometrical frustration. Our results show that the MPS encoding the Schmidt coefficients is weakly entangled even when the entanglement entropy of the decomposed state is strong. This justifies the efficiency of using MPS to encode the Schmidt coefficients, and promises an exponential speedup on the full-state sampling tasks.
\end{abstract}

\maketitle

Despite tremendous successes in the classical simulations of quantum many-body systems achieved by tensor network (TN)~\cite{VMC08MPSPEPSRev, CV09TNSRev, O14TNSRev, RTPC+17TNrev}, there exist severe restrictions concerning the area laws of entanglement entropy (EE)~\cite{ECP10AreaLawRev}. In general, the entanglement scaling of a TN state is determined by the geometric structure of the network, i.e., how the tensors are connected. For instance, the matrix product state (MPS) exhibits a one-dimensional (1D) structure, where the bipartition gives zero-dimensional boundaries [meaning the boundary length satisfies $L_{\partial} \sim O(l^{0})$ with $l$ the length scale]. Consequently, the MPS satisfies the 1D area law of EE and provides a faithful representation of a subclass of states such as the ground states of gapped 1D models with local interactions~\cite{VC06MPSFaithfully, SWVC08MPSent}. 

The projected entangled pair state (PEPS) generalizes MPS to two and higher dimensions~\cite{VC06PEPSArxiv, VWPC06PEPSfamous}. The boundary length of the PEPS with a $D$-dimensional network graph scales as $L_{\partial} \sim O(l^{D-1})$. The simulations of PEPS, including its normalization and the evaluations of EE and observables, concern the contractions of $D$-dimensional TNs, which are usually $\#$P-complete~\cite{SWVC07PEPSAreaLaw, HHZG20PEPSNP}. Particularly, the number of the Schmidt coefficients (i.e., the dimension of the entanglement spectrum) scales exponentially as $\sim O(\chi^{L_{\partial}})$ with $\chi$ the virtual dimension of the TN. The Schmidt decomposition of a TN state can be efficiently done with a constant bipartition length, such as the MPS and tree TN state~\cite{SDV06TTN, O14TNSRev, PhysRevB.105.214201}. \R{For the bipartition whose length scales with system size, matrix product operator  (MPO) was proposed to efficiently access the dominant part of Schmidt coefficients from the tree TN state~\cite{PhysRevB.88.195102}. Valid methods} to access the Schmidt coefficients of the large or infinite-size states with a nontrivial scaling of the boundary length \R{are strongly desired}.

\begin{figure}[tbp]
	\centering
	\includegraphics[angle=0,width=0.95\linewidth]{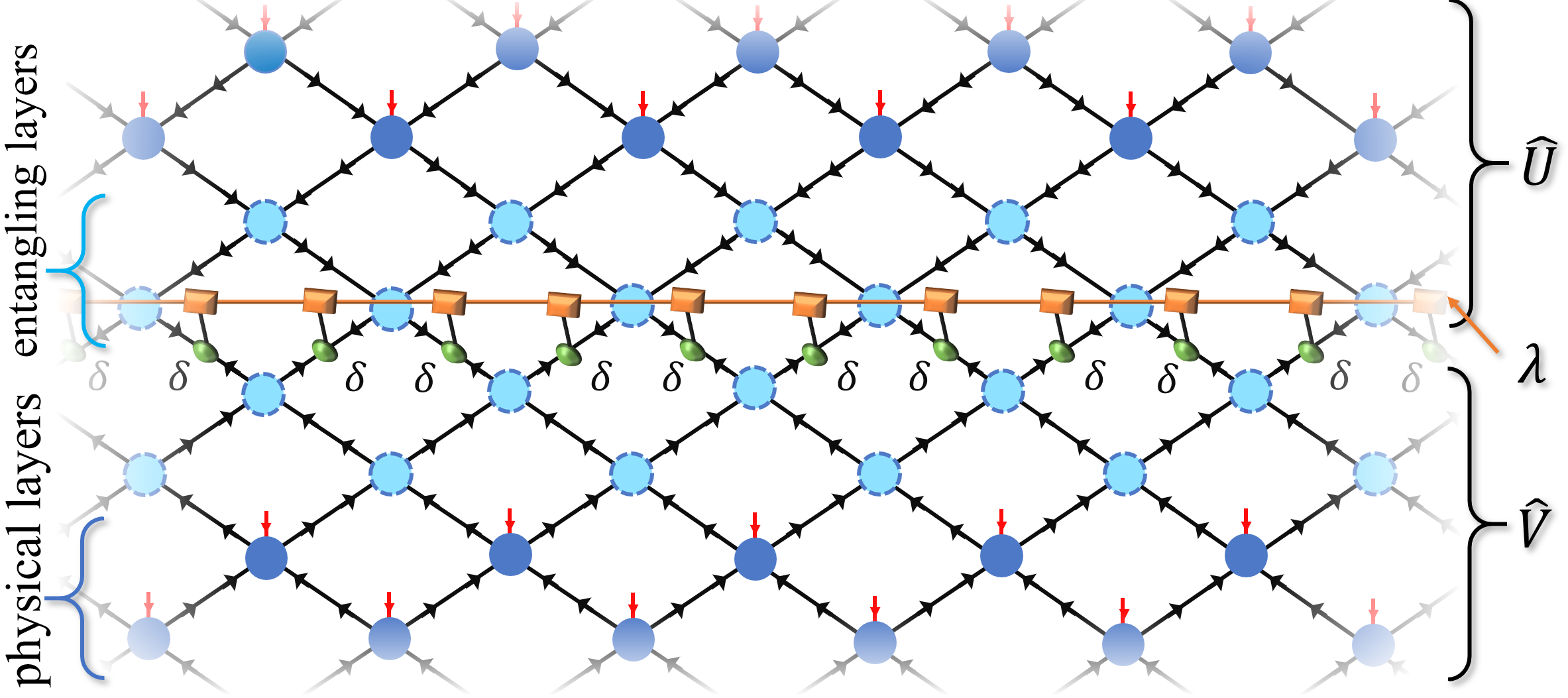}
	\caption{Illustration of the Schmidt TNS formed by two unitary TNs ($\hat{U}$ and $\hat{V}$) and an MPS encoding the Schmidt coefficients $\lambda$ (indicated by the yellow squares). \RR{See the descriptions in detail in the main text}.}
	\label{fig-sTN}
\end{figure}

In this work, we propose a TN state that explicitly involves the Schmidt decomposition of quantum many-body states, which we dub as Schmidt TNS (see the illustration in Fig.~\ref{fig-sTN}). The exponentially-many Schmidt coefficients (also called the entanglement spectrum) are encoded in an MPS, whose complexity is just linear to the number of spins $N$. The transformations in the Schmidt decomposition are represented by the TN's formed by local unitary tensors, where the complexities are also reduced to be linearly with $N$. The Schmidt decomposition for $N \to \infty$ is efficiently reached by imposing the translational invariance on both the unitary TN's and the MPS. 

The power of the Schmidt-TNS on representing the ground state and its Schmidt decomposition is demonstrated by simulating the interacting spin models on a quasi-one-dimensional geometrically frustrated zigzag-pentagon lattice. The weak entanglement of the MPS that encodes the Schmidt coefficients is uncovered, which justifies the validity of MPS to encode the Schmidt coefficients. By using the MPS as a sampler defined in a $2^{L_{\partial}}$-dimensional Hilbert space, an exponential speedup on the full-state sampling is promised by applying the MPS-based schemes~\cite{CPFS+10Tomog, WHWL+20TNtomo}.

\textit{Schmidt tensor network state.---} Considering a quantum state of $N$ spin-$1/2$'s and its bipartition to two subsystems $\mathcal{A}$ and $\mathcal{B}$, the Schmidt decomposition can be written as $ |\Psi \rangle = \sum_{r} \lambda_r |\psi_r\rangle |\phi_r\rangle$ with $\lambda_1 \geq \lambda_2 \geq \ldots \geq 0$ denoting the Schmidt coefficients or entanglement spectrum. $|\psi_r\rangle$ and $|\phi_r\rangle$ are the quantum states defined in the subsystems $\mathcal{A}$ and $\mathcal{B}$, respectively, which we dub as the left and right Schmidt states. They correspond to the left and right singular vectors of the state coefficients $\Psi_{\mathcal{S}_{\mathcal{A}} \mathcal{S}_{\mathcal{B}}}$ with $|\Psi \rangle = \sum_{\mathcal{S}_{\mathcal{A}} \mathcal{S}_{\mathcal{B}}} \Psi_{\mathcal{S}_{\mathcal{A}} \mathcal{S}_{\mathcal{B}}} |\mathcal{S}_{\mathcal{A}} \rangle |\mathcal{S}_{\mathcal{B}} \rangle$ and $\mathcal{S}_{\mathcal{A}(\mathcal{B})} \equiv \prod_{n \in \mathcal{A} (\mathcal{B})} s_{n}$ referring to the spin indices in the subsystem $\mathcal{A}$ (or $\mathcal{B}$). 

The index $r$ can be rewritten in a binary form as $r \equiv (r_1, r_2, \ldots r_{R})$ with $r_m=0, 1$. The number of binary indexes satisfies $R \leq \tilde{R} \equiv \min(\# \mathcal{A}, \# \mathcal{B})$ with $\# \mathcal{A} (\mathcal{B})$ the number of spins in the subsystem $\mathcal{A}$ (or $\mathcal{B}$). The equality holds in the full-rank cases. When the EE satisfies an area law instead of volume law, the number of binary indices for a well-approximated Schmidt decomposition can be compressed to $R \sim O(L_{\partial}) \ll \tilde{R}$. Note it is not difficult to generalize the above discussions to higher-level spins.

The states $|\psi_r\rangle$ and $|\phi_r\rangle$ can be obtained by implementing the unitary transformations on the product state defined by $\{r_1, r_2, \ldots r_{R}\}$ as
\begin{eqnarray}\label{eq-unitary}
	|\psi_r\rangle = \hat{U} \prod_{\otimes m=1}^{R} |r_m\rangle, \ \ |\phi_r\rangle = \hat{V} \prod_{\otimes m=1}^{R} |r_m\rangle.
\end{eqnarray}
The operators $\hat{U}$ and $\hat{V}$ are named as the left and right transformation unitaries, respectively. When $R$ scales linearly with $N$ (say $R = N/2$ with an equal bipartition $\# \mathcal{A} = \# \mathcal{B}$), the complexity of the Schmidt decomposition scales exponentially with $N$. 

\R{Below, we focus on the two-dimensional quantum systems.} In the Schmidt TNS, the Schmidt coefficients are encoded in an MPS $| \lambda \rangle \equiv \sum_{r} \lambda_{r_1 r_2 \ldots r_{R}} \prod_{\otimes m=1}^{R} |r_{m}\rangle$ as 
\begin{eqnarray}\label{eq-MPS}
	| \lambda \rangle = \sum_{r_1r_2 \ldots r_{R}} \text{tTR} \left( \prod_{m=1}^{R} A^{[m]}_{r_m \alpha_m \alpha_{m+1}}\right) \prod_{\otimes m'=1}^{R} |r_{m'}\rangle,
\end{eqnarray}
with $\text{tTR}$ tracing over all shared indices $\{\alpha_m\}$ (which are called the virtual indices of the MPS). The indices $\{r_m\}$ are dubbed as the Schmidt indices of the MPS. Note we do not call $\{r_m\}$ the physical indices according to the MPS terminology to avoid the confusion with the physical indices of the Schmidt TNS that represent the degrees of freedom of the physical spins. The complexity of the MPS scales as $O(Rd_sd_{c}^2)$ (with $d_s$ the dimension of the Schmidt index), which is linear to $R$, while in contrast the dimension of $|\lambda \rangle$ scales exponentially as $O(d_s^R)$. \R{An existing way of lowering the complexity of simulating the Schmidt coefficients is to represent the reduced density matrix as MPO~\cite{PhysRevB.88.195102}. The Schmidt coefficients can be obtained by the density matrix renormalization group (DMRG) to simulate the dominant eigenstates of the MPO~\cite{PhysRevLett.108.067202}. Therefore, one can usually calculate a small number of the dominant Schmidt coefficients. This makes an essential difference from our method, where the MPS $| \lambda \rangle$ here can approximately encode the full exponentially-many Schmidt coefficients.}

The unitary transformations $\hat{U}$ and $\hat{V}$ are represented as the TNs formed by local unitary tensors (see the upper and lower halves of the Schmidt TNS in Fig.~\ref{fig-sTN}). \R{The arrows indicate the orthogonality. By summing the inward indexes of a tensor and its conjugate, one obtains an identity consisting of the outward indexes.} The two unitary TNs ($\hat{U}$ and $\hat{V}$) and the MPS $| \lambda \rangle$ are connected by the third-order superidentical tensors $\delta$ (green circles) that satisfies $\delta_{abc}=1$ if $a=b=c$, or $\delta_{abc}=0$ otherwise). The $\delta$'s map $|\lambda\rangle$ to a diagonal matrix multiplied with $\hat{U}$ and $\hat{V}$ in the expected way required by the Schmidt decomposition. The red indices correspond to the degrees of freedom of quantum spins (called the physical indices in the TN language) and are defined to be inward. The shared black bonds (called the virtual indices) determine the network structure of the unitary TNs. The dimensions of the virtual indices can be set flexibly. 

The tensors in the unitary TNs can be understood as the quantum gates that entangle the product states $\prod_{\otimes m=1}^{R} |r_{m}\rangle$ to the left or right Schmidt states $|\psi_r\rangle$ and $|\phi_r\rangle$. For simplicity, we assume that each tensor may contain zero or one physical index, illustrated by the light or dark blue circles, respectively. A special architecture is shown as an example in Fig.~\ref{fig-sTN}, where we first pretransform $\prod_{\otimes m=1}^{R} |r_{m}\rangle$ to an entangled state by several layers of the tensors without physical index (dubbed as the entangling layers), and then use the tensors with physical indices (dubbed as the physical layers) to map it to one of the Schmidt state. We shall stress that the architecture of the Schmidt TNS, which includes the network structure and the arrangement of the physical/entangling layers, can be flexibly designed for the simulations of different models. \R{More details about the contraction procedures and complexity analyses can be found in the Supplemental Material~\cite{SM}.}

\begin{figure}[tbp]
	\centering
	\includegraphics[angle=0,width=0.9\linewidth]{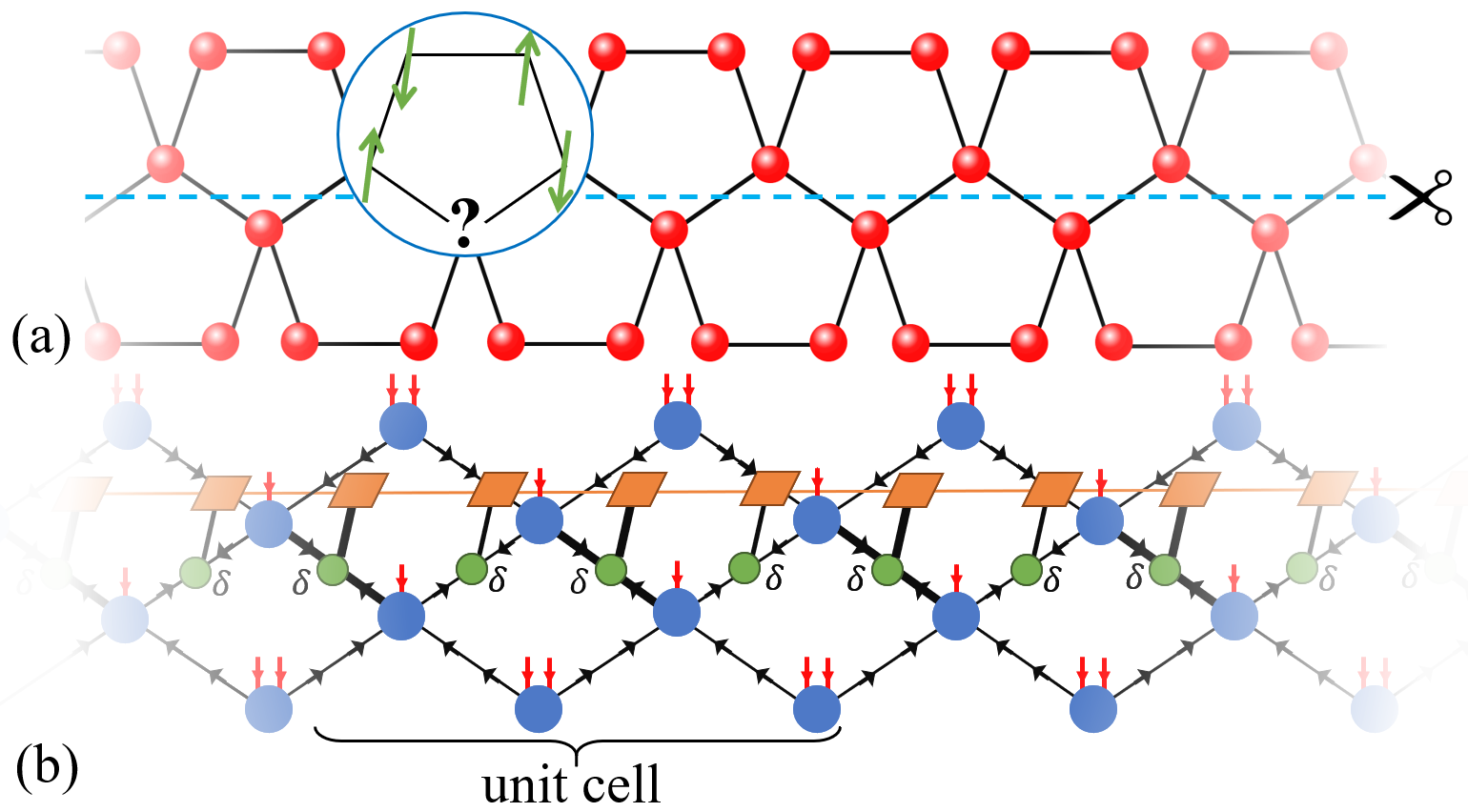}
	\caption{(a) The illustration of the frustrated quasi-one-dimensional ZPAF \RR{and a non-trivial bipartition (dash line). (b) The Schmidt TNS for simulating ZPAF.}}
	\label{fig-lattice}
\end{figure}

\textit{Benchmarks.---} The ground state of a given Hamiltonian $\hat{H}$ can be reached by variationally minimizing the energy $E = \frac{\langle \Psi| \hat{H}| \Psi \rangle}{\langle \lambda|\lambda \rangle}$, where we here take $| \Psi \rangle$ to be a Schmidt TNS. Owing to the unitary property of $\hat{U}$ and $\hat{V}$, the normalization of the Schmidt TNS is equivalent to that of the MPS as $\langle \Psi|\Psi \rangle = \langle \lambda|\lambda \rangle = 1$. The denominator $\langle \lambda|\lambda \rangle$ in $E$ is introduced to manually satisfy the normalization of $| \Psi \rangle$ during the optimization. Each tensor, say $T$, is optimized by the gradient descent as $T \leftarrow T - \eta \frac{\partial E}{\partial T}$ with $\eta$ the gradient step. The gradients can be obtained using the automatic differentiation technique by, e.g., Pytorch~\cite{PyTorch}. To satisfy the unitary conditions for the tensors in $\hat{U}$ and $\hat{V}$, we use the singular vectors of the updated tensors to define the unitary tensors, meaning $T = PSQ^{\dagger}$ by singular value decomposition and then $PQ^{\dagger} \to T$. This trick was originally proposed in the entanglement renormalization to update the disentanglers~\cite{V07EntRenor}, and has been recently applied in the optimization of TNs for quantum computing and machine learning~\cite{LRWP+17MLTN, PRXRITECQC2021, ZHR21ADQC, arxivDMPSshallowCircuit2022}. \R{The ability on mapping product states to target entangled states by local unitary tensors (gates) is crucial to satisfy Eq.~(\ref{eq-unitary}).} The positivity of $|\lambda\rangle$ is guaranteed by mapping the elements of the updated tensors in the MPS to their square.

We consider the spin-$1/2$ zigzag-pentagon antiferromagnet (ZPAF) as an example to demonstrate the validity of the Schmidt TNS [see the illustration in Fig.~\ref{fig-lattice} (a)]. The Hamiltonian can be written as $\hat{H} = \sum_{\langle i,j\rangle} \hat{h}_{i,j}$ with $\hat{h}_{i,j}$ the interaction between the $i$th and $j$th spins, and $\langle i,j\rangle$ denotes the nearest-neighbor spin pairs (marked by the black lines). We consider the antiferromagnetic Heisenberg interactions with $\hat{h}_{ij} = \sum_{\alpha=x, y, z} \hat{S}_i^{\alpha} \hat{S}_j^{\alpha}$ and the $XY$ interactions $\hat{h}_{ij} = \sum_{\alpha=x, y} \hat{S}_i^{\alpha} \hat{S}_j^{\alpha}$, with $\hat{S}_i^{\alpha}$ the spin operator on the $i$th site in the $\alpha$ direction. The ZPAF is geometrically frustrated since there is no arrangement where each nearest-neighboring spin pair is aligned antiparallelly. Geometrical frustration could usually lead to a macroscopic ground-state degeneracy that melts the magnetic orders even at zero temperature and meanwhile induce strong entanglement~\cite{MR06FrustrateRev}. The ground state of ZPAF exhibits a vanishing average magnetization $M \simeq O(10^{-5})$.

For a quasi-1D system, the DMRG requires a 1D path to define the MPS~\cite{SW12DMRG2DRev}. This path should go through all sites. The Schmidt decomposition can be accurately reached using the orthogonal forms or in the infinite cases canonical form of the MPS~\cite{OV08canonical, S11DMRGRev}. However, the bipartition boundary has to be a zero dimensional, meaning $L_{\partial} \sim O(l^0)$. The Schmidt TNS allows one to achieve the Schmidt decomposition with the bipartition along the stretching direction of the quasi-1D systems, where the length of the bipartition boundary scales linearly with the system size as $L_{\partial} \sim O(N)$ (see the horizontal blue dashed line). This cannot be realized with the existing methods.

The architecture of the Schmidt TNS for ZPAF is designed in such a way that for each tensor in $\hat{U}$ and $\hat{V}$, the total dimension of the inward indices equals to that of the outward indices [see Fig.~\ref{fig-lattice} (b) and the figure caption for details]. Note this is not a mandatory requirement, but can keep the tensors to be unitary instead of isometric. A direct consequence is that all Schmidt coefficients will be kept (with $R = \tilde{R}$). Note that isometries can be flexibly introduced in the TNs of $\hat{U}$ and $\hat{V}$ to compress the number of Schmidt coefficients (i.e., the dimension of $|\lambda\rangle$). 

\begin{figure}[tbp]
	\centering
	\includegraphics[angle=0,width=0.85\linewidth]{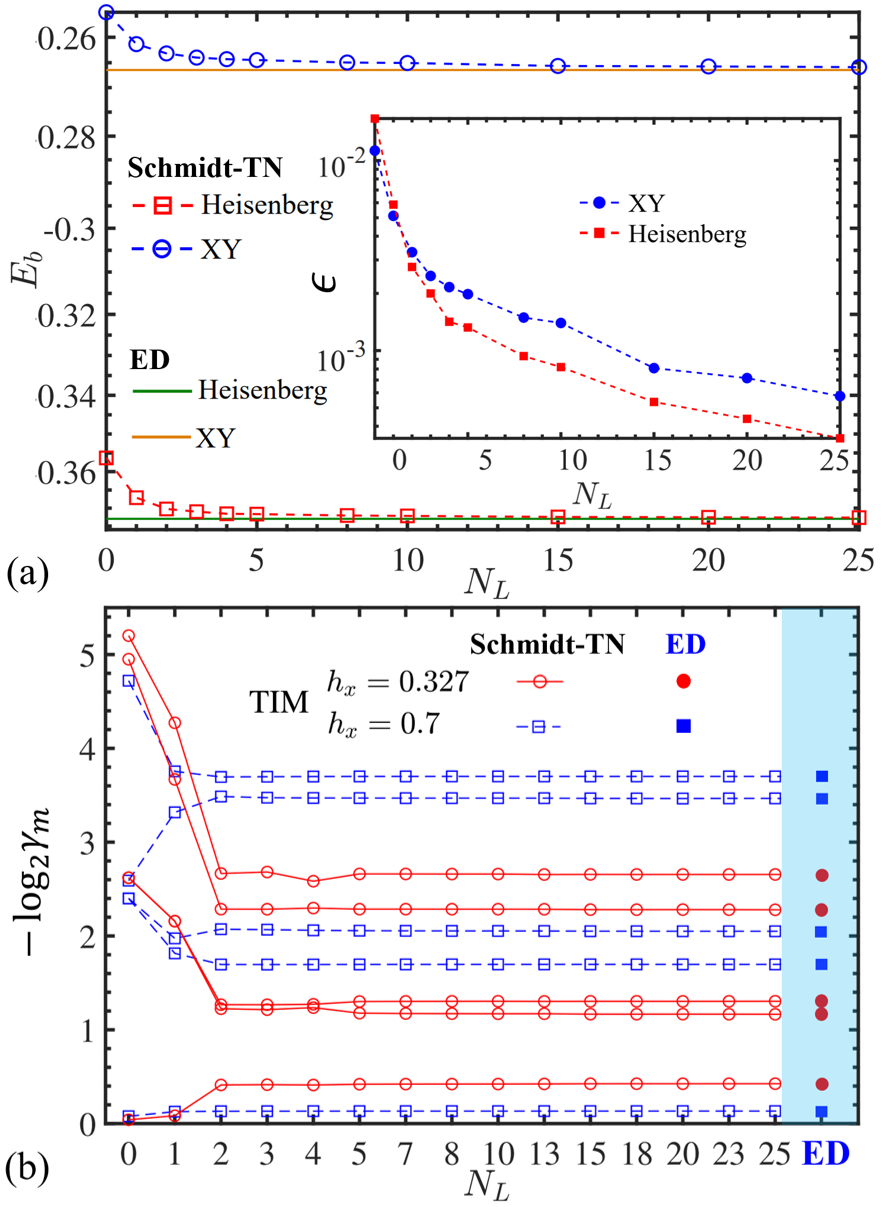}
	\caption{(a) \RR{The ground-state energy density $E_b$ ($N=16$) with the Heisenberg and XY interactions versus the number of entangling layers $N_L$. The inset shows the error $\epsilon$.} (b) The negative logarithm of the five largest Schmidt coefficients ($-\log_2 \gamma_{m}$) of the ground state with the Ising interaction in a transverse field $h_x$ (TIM). \RR{The Schmidt coefficients obtained by ED are shown in the blue-shadowed column.}}
	\label{fig-finiteE}
\end{figure}

Figure.~\ref{fig-finiteE} (a) shows the ground-state energy density (i.e., per nearest-neighboring pair) $E_b$ by varying the number of entangling layers $N_L$ (with $N=16$). The inset shows the error $\epsilon$ by comparing with the exact diagonalization (ED). We take the virtual bond dimensions to be sufficiently large; thus the error is mainly controlled by $N_L$. This parameter controls how well $\hat{U}$ and $\hat{V}$ are reached by local unitaries. The validity of the Schmidt TNS is supported by the exponentially decreasing $\epsilon$ versus $N_L$. Our results are consistent with the previous works in quantum computation showing that a large class of states can be efficiently reached by the circuits with local unitary gates~\cite{CPFS+10Tomog, R20MPSencode, ZHR21ADQC, ZYY21StatePre, GL22MPSPre, arxivDMPSshallowCircuit2022}. 

\begin{table*}[tbp]
 \setlength\tabcolsep{4pt} 
 \renewcommand{\arraystretch}{1.0}
 \begin{center}
 \end{center}
 \centering          
 \begin{tabular*}{0.84\linewidth}{c|c|cc|ccc}
 	\hline \hline 
 	$E_b$ & \makecell[c]{ED} &\multicolumn{2}{c}{\makecell[c]{DMRG}}&\multicolumn{3}{|c}{\makecell[c]{iSchmidt-TNS, $N=\infty$}}\\ \hline
 	\makecell[c]{\diagbox{model}{parameters}} & \makecell[c]{$N=18$} & $N=18$&$N=180$&$N_L=0$&$N_L=1$&$N_L=2$\\ \hline
 	TIM, $h_x$=0.5&-0.2767646 &-0.2767646&-0.2517436 & -0.2416401 &-0.2550272 &-0.2556101 \\ \hline
 	TIM, $h_x$=0.2&-0.2120327&-0.2120327&-0.1966008 & -0.1950901 & -0.1988530 & -0.1988552 \\ \hline
 	$XY$ &-0.2732445&-0.2732445&-0.2527285& -0.2350790 & -0.2545480 &-0.2572155 \\ \hline
 	Heisenberg&-0.3842125&-0.3842125&-0.3522980& -0.3319882 &-0.3589855 &-0.3607915 \\ \hline
 	\hline
 \end{tabular*}               
 \caption{The ground-state energy density $E_{b}$ of the ZPAF with the Heisenberg, $XY$, or Ising interactions in a transverse field (TIM). The $E_{b}$ of infinite size is given by the iSchmidt TNS with different numbers of the entangling layers $N_{L}$ (with the MPS virtual dimension $\chi=2$). The results with $N=18$ by ED and DMRG, and those with $N=180$ by DMRG (dimension cutoff $\chi_{c}=200$) are given for comparison.}
 \label{tab}
\end{table*}

In Fig.~\ref{fig-finiteE} (b), we consider the transverse Ising mode  (TIM) with the Hamiltonian $\hat{H} = \sum_{\langle i,j\rangle} \hat{S}^{z}_{i} \hat{S}^{z}_{j} - h_{x} \sum_{n} \hat{S}^{x}_{n}$. We take $h_{x} = 0.327$ where the ground-state EE reaches its maximal with $S = - 2\sum_{m} \gamma_{m}^{2} \log_{2} \gamma_{m} \simeq 1.25 $ (with $\gamma_{m}$ the $m$th Schmidt coefficient), and $h_{x} = 0.7$ where the system is in a much less entangled polarized phase with $S \simeq 0.63$. The negative logarithms of the five largest Schmidt coefficients are displayed with various $N_{L}$. The Schmidt TNS accurately gives the Schmidt coefficients in both cases for about $N_{L} > 4$, compared with those from the Schmidt decomposition of the full ground states obtained by ED. 

For the infinite-size ZPAF, the horizontal cut will lead to an infinitely long bipartition boundary. There is currently no valid method to access the Schmidt decomposition that contains infinitely many coefficients. Table \ref{tab} shows the ground-state energy density $E_{b}$ of the infinite-size ZPAF with the Heisenberg, $XY$, or Ising interactions in a transverse field obtained by the infinite Schmidt TNS imposed with translational invariance (iSchmidt TNS for short). We assume that each unit cell contains twelve inequivalent tensors, of which eight from the unitary TNs and four from the MPS [see Fig.~\ref{fig-lattice} (b)]. 

The results from ED and DMRG~\cite{W92DMRG, W93DMRG} are given for comparison. When the system size is small (say $N=18$), the approximation error in DMRG is ignorable. The results from DMRG and ED are almost identical. The finite-size effects are dominative, and consequently its ground-state energy is generally much lower than the true ground-state energy in the thermodynamic limit (denoted as $E_{\infty}$). As the size increases to, e.g., $N=180$, the finite-size effects become insignificant. The error of DMRG mainly comes from the approximations. The obtained energy should give an upper bound of $E_{\infty}$. The dimension cutoff in DMRG is taken as $\chi_{c}=200$ where $E_{b}$ already converges.

For the iSchmidt TNS, the obtained energy should also be an upper bound, since the normalization of the iSchmidt TNS is strictly kept by the division by $\langle \lambda| \lambda \rangle$. With no entangling layer ($N_L = 0$), the energy of the iSchmidt TNS is much higher than that with $N=180$ by DMRG. By introducing $N_{L}=1$ entangling layer, the iSchmidt TNS reaches a better lower bound than DMRG. By increasing to $N_{L} = 2$ entangling layers, the energy converges with a slight change of O$(10^{-3})$. These indicate the $E_{\infty}$ is accurately reached by the iSchmidt-TNS with an error about O$(10^{-3})$.

\begin{table}[tbp]
 \setlength\tabcolsep{4pt} 
 \renewcommand{\arraystretch}{1.0}
 \begin{center}
 \end{center}
 \centering          
 \begin{tabular*}{1\linewidth}{c|ccc}
 	\hline \hline 
 	\makecell[c]{$E_{b}$} & \makecell[c]{$\chi =1$} &$\chi =3$&$\chi =5$\\ \hline
 	TIM, $h_x$=0.5 & -0.25499767 & -0.25503064 & -0.25503065\\ \hline
 	TIM, $h_x$=0.2 & -0.19883578 & -0.198853221 & -0.19885543 \\ \hline
 	XY & -0.25449082 & -0.25455178 & -0.25455190 \\ \hline
 	Heisenberg & -0.35876465 & -0.35903425 & -0.35903479 \\ \hline
 	\hline
 \end{tabular*}                 
 \caption{The ground-state energy density $E_{b}$ of the infinite-size ZPAF with the Heisenberg, XY, or Ising interactions in a transverse field (TIM) given by the iSchmidt TNS with different virtual dimensions $\chi$ of the MPS $| \lambda \rangle$. We fix $N_L=1$. }
 \label{tab-mpsent}
\end{table}

\textit{Acceleration on full-state sampling.---} The complexity of the full-state sampling on $N$ spins scales exponentially as O$(2^{N})$. The MPS $|\lambda\rangle$ in the Schmidt TNS is normalized; thus it can be treated as a quantum state of $R$ spins $\{|r_{m} \rangle\}$ ($m=1, \ldots, R$). The probability satisfies $P(r_1, r_2, \ldots r_{R}) = \left|\langle \lambda | \prod_{\otimes m=1}^{R} |r_{m}\rangle \right|^{2}$. The dimension of the sampling space is smaller than that of the quantum state $ |\Psi \rangle$. In our example with the equal bipartition, acceleration can be gained on the full-state sampling since we have $\text{dim}(|\lambda\rangle) = \sqrt{\text{dim} (|\Psi \rangle)}$.

More notably, the Schmidt coefficients can be efficiently encoded into a weakly entangled state. In Table~\ref{tab-mpsent}, we show the ground-state energy of ZPAF with different virtual dimension $\chi$ of the MPS. The entanglement entropy of the MPS (not the state $|\Psi \rangle$) satisfies $S_{\text{MPS}} \leq \ln{\chi}$. For $\chi=1$, the MPS gives a product state with no entanglement ($S_{\text{MPS}} = 0$). The difference of the ground-state energy by increasing to $\chi=5$ is just O$(10^{-4}) \sim$ O$(10^{-5})$. These results suggest that the MPS representation with a small bond dimension is sufficient to encode the Schmidt coefficients. Compared with the direct state sampling on $|\Psi\rangle$, an exponential speedup is promised by using the efficient MPS-based schemes for sampling on $|\lambda \rangle$, where the complexity scales just linearly with $R$~\cite{CPFS+10Tomog, WHWL+20TNtomo}. 

\textit{Summary.---} We propose the Schmidt TNS that efficiently represents the Schmidt decomposition of the quantum many-body state with linearly scaled complexity. The transformations in the Schmidt decomposition are given by two TNs consisting of local unitaries, and the Schmidt coefficients are encoded in a matrix product state. The Schmidt decomposition of infinite-size states can be reached by imposing translational invariance. The validity of Schmidt TNS is demonstrated by simulating the ground state of a quasi-one-dimensional frustrated spin model. \RR{Without truncations, the computational cost scales exponentially with either the width of the lattice or the depth of the unitary TN's, which can be lowered by incorporating with other TN contraction algorithms (e.g.,~[\onlinecite{V03TEBD, V04TEBD, JOVVC08PEPS}].} The states encoding the Schmidt coefficients are weakly entangled, thus can be well represented as an MPS with a small bond dimension. This promises a significant acceleration of the full-state sampling using the MPS-based methods.

 \section*{Acknowledgment}
P.F.Z is thankful to Ding-Zu Wang, Wei-Ming Li, Pei Shi, and Xiao-Han Wang  for stimulating discussions. This work was supported in part by NSFC (Grants No. 12004266 and No. 11834014), Beijing Natural Science Foundation (Grant No. 1232025), and Academy for Multidisciplinary Studies, Capital Normal University.

%

\end{document}